# AN INTERFACE CRACK IN 1D PIEZOELECTRIC QUASICRYSTAL UNDER ANTIPLANE MECHANICAL LOADING AND ELECTRIC FIELD

Altoumaimi M., Loboda V.V.

***Abstract.*** *The present study provides the consideration of a mode III interface crack in one-dimentional (1D) piezoelectric quasicrystal under antiplane phonon and phason loading and inplane electric field. Due to complex function approach all required electromechanical parameters are presented through vector-functions analytic in the whole complex plane except the crack region.*

*The cases of electrically impermeable (insulated) and electrically limited permeable conditions on the crack faces are considered. In the first case a vector Hilbert problem in the complex plane is formulated and solved exactly and in the second one the quadratic equation with respect to the electric flux through the crack region is obtained additionally. Its solution permits to find phonon and phason stresses, displacement jumps (sliding) and also electric characteristics along the material interface. Analytical formulas are also obtained for the corresponding stress intensity factors related to each field.*

*The numerical computations for three selected variants of the loading conditions was conducted and the resulting field distributions are visualised on the crack continuation beyond the crack and also inside of the crack region.*

## 1    INTRODUCTION

Quasicrystals (QCs) are distinguished by their long-range orientation order and quasiperiodic translational symmetry, endowing them with exceptional mechanical and functional properties compared to conventional crystalline materials. The seminal discovery by Shechtman *et al.* (1984) of metallic alloys exhibiting non-periodic long-range order and high hardness stimulated extensive investigation into the elasticity and fracture behaviour of these materials [1].

The foundational continuum theory of QC elasticity incorporates coupled phonon and phason displacement fields. In pioneering work, Ding *et al.* (1993) formulated the generalized elastic constitutive relations for quasicrystals, deriving the full electro-elastic coupling between phonon and phason modes [2]. Building on this framework, Fan (2011) provided a comprehensive mathematical treatment of QC elasticity and its applications, including explicit solutions for fundamental boundary-value problems [3]. Fracture mechanics of QCs under anti-plane shear (Mode III) has been addressed in several studies. Shi *et al.* (2007) analysed interfacial cracks between conventional elastic materials and quasicrystals, highlighting the role of phason fields in crack tip stress singularities [4]. Zhou and Li (2018) derived exact solutions for two collinear cracks normal to a 1D hexagonal piezoelectric QC boundary, obtaining closed-form expressions for stress intensity factors and displacement fields [5].

A non-uniformly loaded anti-plane crack embedded in a half-space of a one-dimensional piezoelectric quasicrystal was studied in [6] and two collinear electrically permeable anti-plane cracks of equal length lying at the mid-plane of a one-dimensional hexagonal piezoelectric quasicrystal strip were investigated in [7]. Two thin strips with a microcrack at the interface were studied in paper [8]. Piezoelectric coupling in QCs introduces additional complexity. Hu *et al.* (2019) reduced the mixed electro-mechanical boundary-value problem for an interface crack in dissimilar 1D hexagonal piezoelectric QCs to singular integral equations via Riemann–Hilbert methods, yielding full-field solutions for phonon, phason, and electric quantities along the crack faces [9]. Govorukha and Kamlah (2024) extended these results by considering mixed electric boundary conditions-combining conducting and permeable crack face segments- demonstrating how partial electrical contact modulates crack-tip intensity factors [10]. Loboda et al. (2022) further generalized to multiple collinear interface cracks in layered piezoelectric QCs, revealing interaction effects on stress intensity factors and energy release rates under coupled electromechanical loading [11].

Electrically limited permeable interface crack model was suggested by Hao and Shen [12] concerning the plane problem for a homogeneous piezoelectric material. Till nowadays an exact anti-plane analytical solution for an electrically limited permeable interface crack between dissimilar piezoelectric QCs, remains to be presented. The present study addresses this gap by (i) formulating the coupled phonon–phason–electric field equations for a bimaterial QC plate, (ii) reducing the interface crack boundary-value problem to a vector Hilbert problem, (iii) deriving closed-form expressions for crack-face opening displacement and electric potential jump, and (iv) obtaining analytical formulas for the stress intensity factors associated with each field. Numerical validation and visualization of field distributions in the crack-continuation region complete the solution, offering a practical tool for design and optimization of QC-based electromechanical systems.

## 2   FORMULATION OF THE BASIC RELATIONS

For the linear elastic theory of QCs, the constitutive relations, equilibrium equations and geometric equations of 1D piezoelectric hexagonal QC with point group 6 $mm$ without body forces and free charges can be expressed in the following form [13]

$$\sigma_{ij} = c_{ijks}\varepsilon_{ks} - e_{sij}E_s + R_{ij3s}w_{3s}, \tag{1}$$

$$D_i = e_{iks}\varepsilon_{ks} + \xi_{is}E_s + \hat{e}_{i3s}w_{3s}, \tag{2}$$

$$H_{3i} = R_{ks3i}\varepsilon_{ks} - \hat{e}_{s3i}E_s + K_{3i3s}w_{3s}, \tag{3}$$

$$\sigma_{ij,j} = 0, \; D_{i,i} = 0, \; H_{3i,i} = 0, \tag{4}$$

$$\varepsilon_{ij} = \tfrac{1}{2}(u_{i,j} + u_{j,i}), \; E_i = -\phi_{,i}, \; w_{3i} = w_{3,i}, \tag{5}$$

where $i, j, k, s = 1, 2, 3$, and the denotation "," represents the derivative operation for the space variables; $u_i$, $w_3$ and $\varphi$ are the phonon displacements, phason displacement, and electric potential, respectively, and the atom arrangement is periodic in the $x_1 - x_2$ plane and quasi-periodic in the $x_3$-axis; $\sigma_{ij}$ and $\varepsilon_{ks}$ are the phonon stresses and strains, respectively; $H_{3i}$ and $w_{3i}$ are the phason stresses and strains, respectively; $D_i$ and $E_i$ are the electric displacements and electric fields, respectively, and the polarization direction is along the $x_3$-axis; $c_{ijks}$ and $K_{3j3s}$ are the elastic constants in the phonon and phason fields, respectively; $R_{ij3k}$ represent the phonon–phason coupling elastic constants; $e_{jks}$ and $\hat{e}_{jks}$ are the piezoelectric constants in the phonon and phason fields, respectively; $\xi_{is}$ are the permittivity constants. Here, a comma in subscript denotes differentiation with respect to the following spatial variable.

For the case of antiplane mechanical loading and an in-plane electric loading with reference to the $x_1 O x_2$-plane all fields are independent of the variable $x_3$. Therefore, the problem under consideration is a so-called anti-plane shear problem or mode-III crack problem. In this case

$$u_1 = u_2 = 0, \; u_3 = u_3(x_1, x_2), \; w_3 = w_3(x_1, x_2), \; \varphi = \varphi(x_1, x_2), \tag{6}$$

and the constitutive relations take the form:

$$\begin{Bmatrix} \sigma_{j3} \\ H_{j3} \\ D_j \end{Bmatrix} = \mathbf{R} \begin{Bmatrix} u_{3,j} \\ w_{3,j} \\ \varphi_{,j} \end{Bmatrix} \quad (j=1,2), \tag{7}$$

Where
$$\mathbf{R} = \begin{bmatrix} c_{44} & R_3 & e_{15} \\ R_3 & K_2 & \hat{e}_{15} \\ e_{15} & \hat{e}_{15} & -\xi_{11} \end{bmatrix}, \tag{8}$$

and $c_{44}, K_2, R_3$ stand for the phonon elastic modulus, phason elastic modulus and phonon-phason coupling modulus, respectively; which are written in the simplified index notation. Also $e_{15}$, $\hat{e}_{15}$ are piezoelectric constants of phonon and phason fields and $\xi_{11}$ is the permittivity. Introducing the vectors

$$\boldsymbol{u} = [u_3, w_3, \varphi]^T, \quad \boldsymbol{t}_j = [\sigma_{3j}, H_{3j}, D_j]^T, \tag{9}$$

one can write
$$\boldsymbol{t}_j = \boldsymbol{R}\boldsymbol{u}_{,j} \quad (j=1,2). \tag{10}$$

For the considered anti-plane problem the equilibrium equations (4) takes the form
$$\frac{\partial \sigma_{31}}{\partial x_1} + \frac{\partial \sigma_{32}}{\partial x_2} = 0, \quad \frac{\partial D_1}{\partial x_1} + \frac{\partial D_2}{\partial x_2} = 0, \quad \frac{\partial H_{31}}{\partial x_1} + \frac{\partial H_{32}}{\partial x_2} = 0.$$

Substituting (7) in the last equation we get that the functions $u_3$, $\varphi$ and $w_3$ satisfy the equations $\Delta u_3 = 0$, $\Delta \varphi = 0$, $\Delta w_3 = 0$, respectively, i.e. they are harmonic. Therefore, present the vector $\boldsymbol{u}$, composed of these function, as real parts of some analytic vector-function
$$\boldsymbol{u} = 2\operatorname{Re}\boldsymbol{\Phi}(z) = \boldsymbol{\Phi}(z) + \overline{\boldsymbol{\Phi}(z)} \quad (2 \text{ is introduced for convenience}), \tag{11}$$

where $\boldsymbol{\Phi}(z) = [\Phi_1(z), \Phi_2(z), \Phi_3(z)]^T$ is an arbitrary analytic function of the complex variable $z = x_1 + ix_2$.

Substituting (11) in (10), one gets
$$\boldsymbol{t}_1 = -i\boldsymbol{B}\boldsymbol{\Phi}'(z) + i\overline{\boldsymbol{B}}\overline{\boldsymbol{\Phi}'(z)}, \quad \boldsymbol{t}_2 = \boldsymbol{B}\boldsymbol{\Phi}'(z) + \overline{\boldsymbol{B}}\overline{\boldsymbol{\Phi}'(z)}, \tag{12}$$

where $\boldsymbol{B} = i\boldsymbol{R}$.

**Bimaterial plane.** Suppose that the plane $(x_1, x_2)$ is composed of two half-planes $x_2 > 0$ and $x_2 < 0$. Different cracks, inclusions and other defects can take place on the axis $x_1$. The presentation (11), (12) can be written for regions $x_2 > 0$ and $x_2 < 0$ which in this case takes the form
$$\boldsymbol{u}^{(m)} = \boldsymbol{\Phi}^{(m)}(z) + \overline{\boldsymbol{\Phi}^{(m)}(z)}, \quad \boldsymbol{t}_2^{(m)} = \boldsymbol{B}^{(m)}\boldsymbol{\Phi}'^{(m)}(z) + \overline{\boldsymbol{B}}^{(m)}\overline{\boldsymbol{\Phi}'^{(m)}(z)}, \tag{13}$$

where $m=1$ for the area 1 and $m=2$ for the area 2; $\boldsymbol{B}^{(m)}$ are the matrixes $\boldsymbol{B}$ for the areas 1 and 2, respectively; $\boldsymbol{\Phi}^{(m)}(z)$ are the arbitrary vector functions, analytic in the areas 1 and 2, respectively.

Next we require that the equality $\boldsymbol{t}_2^{(1)} = \boldsymbol{t}_2^{(2)}$ holds true on the entire axis $x_1$. Then it follows from (13)
$$\boldsymbol{B}^{(1)}\boldsymbol{\Phi}'^{(1)}(x_1 + i0) + \overline{\boldsymbol{B}}^{(1)}\overline{\boldsymbol{\Phi}'^{(1)}}(x_1 - i0) = \boldsymbol{B}^{(2)}\boldsymbol{\Phi}'^{(2)}(x_1 - i0) + \overline{\boldsymbol{B}}^{(2)}\overline{\boldsymbol{\Phi}'^{(2)}}(x_1 + i0). \tag{14}$$

Here we have used the first form of designation $F(x_1 \pm i0) = F^{\pm}(x_1)$, which refers to the limit value of a function $F(z)$ at $y \to 0$ from above or below, respectively.

The equation (14) can be written as

$$B^{(1)}\Phi'^{(1)}(x_1+i0)-\bar{B}^{(2)}\bar{\Phi}'^{(2)}(x_1+i0)=B^{(2)}\Phi'^{(2)}(x_1-i0)-\bar{B}^{(1)}\bar{\Phi}'^{(1)}(x_1-i0).$$

The left and right sides of the last equation can be considered as the boundary values of the functions
$$B^{(1)}\Phi'^{(1)}(z)-\bar{B}^{(2)}\bar{\Phi}'^{(2)}(z) \text{ and } B^{(2)}\Phi'^{(2)}(z)-\bar{B}^{(1)}\bar{\Phi}'^{(1)}(z), \tag{15}$$
which are analytic in the upper and lower planes, respectively. But it means that there is a function $M(z)$, which is equal to the mentioned functions in each half-plane and is analytic in the entire plane. This function is the following:

$$M(z)=\begin{cases} B^{(1)}\Phi'^{(1)}(z)-\bar{B}^{(2)}\bar{\Phi}'^{(2)}(z) & \text{for } x_2>0 \\ B^{(2)}\Phi'^{(2)}(z)-\bar{B}^{(1)}\bar{\Phi}'^{(1)}(z) & \text{for } x_2<0 \end{cases}$$

Assuming that $M(z)|_{z\to\infty}\to 0$, on the basis of the Liouville theorem we find that each of the functions (15) is equal to 0 for each $z$ from the corresponding half-plane. Hence, we obtain
$$\bar{\Phi}'^{(2)}(z)=\left(\bar{B}^{(2)}\right)^{-1}B^{(1)}\Phi'^{(1)}(z) \text{ for } x_2>0, \tag{16}$$
$$\bar{\Phi}'^{(1)}(z)=\left(\bar{B}^{(1)}\right)^{-1}B^{(2)}\Phi'^{(2)}(z) \text{ for } x_2<0. \tag{17}$$

Further we find the jump of the following vector function
$$\langle u'(x_1)\rangle=u'^{(1)}(x_1+i0)-u'^{(2)}(x_1-i0), \tag{18}$$
when passing through the interface. Finding from the first formula (13)
$$u'^{(m)}(z)=\Phi'^{(m)}(z)+\bar{\Phi}'^{(m)}(\bar{z})$$
or
$$u'^{(m)}(x_1\pm i0)=\Phi'^{(m)}(x_1\pm i0)+\bar{\Phi}'^{(m)}(x_1\mp i0)$$
and substituting in (18), one gets
$$\langle u'(x_1)\rangle=\Phi'^{(1)}(x_1+i0)+\bar{\Phi}'^{(1)}(x_1-i0)-\Phi'^{(2)}(x_1-i0)-\bar{\Phi}'^{(2)}(x_1+i0).$$

Finding further of (17) $\Phi'^{(2)}(x_1-i0)=\left(B^{(2)}\right)^{-1}\bar{B}^{(1)}\bar{\Phi}'^{(1)}(x_1-i0)$ and substituting this expression together with (16), at $x_2\to +0$, in the latest formula, leads to
$$\langle u'(x_1)\rangle=D\Phi'^{(1)}(x_1+i0)+\bar{D}\bar{\Phi}'^{(1)}(x_1-i0),$$
where $D=I-\left(\bar{B}^{(2)}\right)^{-1}B^{(1)}$, $I=diag[1,1,1]$ – the identity matrix.

Introducing a new vector-function
$$W(z)=\begin{cases} D\Phi'^{(1)}(z), & x_2>0 \\ -\overline{D\Phi'^{(1)}}(z), & x_2<0 \end{cases}, \tag{19}$$
the expression for the derivative of displacement jump can be written as
$$\langle u'(x_1)\rangle=W^+(x_1)-W^-(x_1). \tag{20}$$

From the second relations (13) we have
$$t_2^{(1)}(x_1,0)=B^{(1)}\Phi'^{(1)}(x_1+i0)+\bar{B}^{(1)}\bar{\Phi}'^{(1)}(x_1-i0). \tag{21}$$

Given that on the base of (19)
$$\Phi'^{(1)}(x_1+i0)=D^{-1}W(x_1+i0),$$
$$\bar{\Phi}'^{(1)}(x_1-i0)=-\left(\bar{D}^{-1}\right)^{-1}W(x_1-i0)$$
and substituting these relations in (21), leads to
$$t_2^{(1)}(x_1,0)=GW^+(x_1)-\bar{G}W^-(x_1), \tag{22}$$

where $G = B^{(1)}D^{-1}$. Simple calculations show that
$$G = \left[\left(B^{(1)}\right)^{-1} + \left(B^{(2)}\right)^{-1}\right]^{-1}, \qquad (23)$$
moreover, $\bar{G} = -G$. Then the equation (22) takes the form
$$t_2^{(1)}(x_1, 0) = G(W^+(x_1) + W^-(x_1)). \qquad (24)$$

The representations (20) and (24) are very convenient for solving of antiplane problems for bimaterial piezoelectric QCs with cracks and inclusions at the interface.

## 3. An-electrically insulated crack between two piezoelectric quasicrystals

Assume that there is a crack $|x_1| < a$ with mechanically free and electrically-insulated faces at the interface $x_2 = 0$ (fig. 1).

The shear phonon $\sigma_{32}^\infty$ and phason $H_{32}^\infty$ stresses and electric flux are prescribed at infinity and are designated by the vector $t_2^\infty = \left[\sigma_{32}^\infty, H_{32}^\infty, D_2^\infty\right]$.

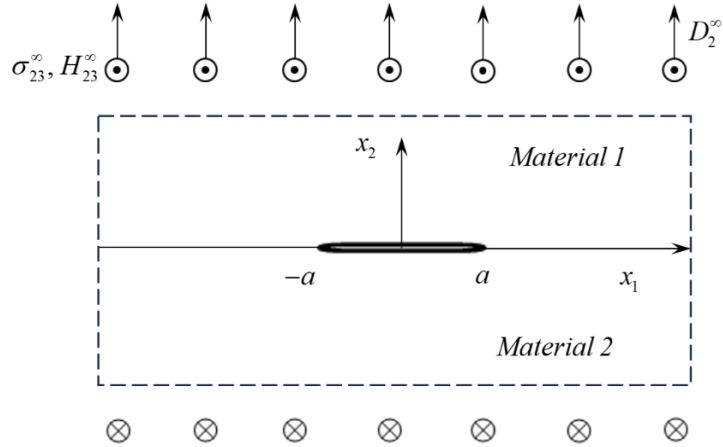

Fig. 1. The crack between two piezoelectric quasicrystals

The boundary conditions at the interface are of the form
$$\sigma_{32}^{(1)} = \sigma_{32}^{(2)} = 0,\ D_2^{(1)} = D_2^{(2)} = 0,\ H_{32}^{(1)} = H_{32}^{(2)} = 0 \text{ for } |x_1| < a, \qquad (25)$$
$$\langle\sigma_{32}\rangle = 0, \langle D_2\rangle = 0, \langle H_{32}\rangle = 0, \langle\varepsilon_{31}\rangle = 0, \langle E_1\rangle = 0,\ \langle w_{31}\rangle = 0 \text{ for } |x_1| > a, \qquad (26)$$
where $\langle * \rangle$ means a jump of the corresponding function when passing through the axis $x_1$.

The first three conditions (26) are satisfied on the basis of (14). From the satisfaction of the last three conditions (26) the analyticity of the vector function $W(z)$ in the whole plane, except of the crack area, follows due to (20). Also the fulfillment of the conditions (25) using (24) lead to the vector Hilbert problem of linear relationship
$$W^+(x_1) + W^-(x_1) = 0 \text{ for } x_1 \in (-a, a). \qquad (27)$$

Taking into account that $W^+(x_1) = W^-(x_1) = W(x_1)$ for $x_1 \notin (-a, a)$ and $W(x_1)|_{x_1 \to \infty} = W(z)|_{z \to \infty}$ we get by using (24) the following condition at infinity
$$W(z)|_{z \to \infty} = \frac{1}{2}G^{-1}t_2^\infty. \qquad (28)$$

The solution of the problem (27) under the condition at infinity (28) has the form [14]

$$W(z) = \frac{1}{2} G^{-1} t_2^\infty \frac{z}{\sqrt{z^2 - a^2}}. \tag{29}$$

Substituting (29) in (24) we obtain:

$$t_2^{(1)}(x_1, 0) = t_2^\infty \frac{x_1}{\sqrt{x_1^2 - a^2}} \quad \text{for } x_1 \notin [-a, a]. \tag{30}$$

Considering that $W^-(x_1) = -W^+(x_1) = W(x_1)$ for $x_1 \in (-a, a)$, one gets from relations (20), (29)

$$\langle \mathbf{u}'(x_1) \rangle = G^{-1} t_2^\infty \frac{x_1}{i\sqrt{a^2 - x_1^2}} \quad \text{for } x_1 \in (-a, a). \tag{31}$$

Integrating of the last relation leads to the following formula for the jump of displacement of the crack faces

$$\langle \mathbf{u}(x_1) \rangle = i G^{-1} t_2^\infty \sqrt{a^2 - x_1^2}. \tag{32}$$

### 4. The case of prescribed electrical displacements at the crack faces

Let's assume that nonzero electric displacement $D_2^{(1)} = D_2^{(2)} = D_0$ is prescribed at the crack faces. In this case one has in the crack region:

$$t_2^{(1)}(x_1, 0) = t_2^{(0)} \quad \text{for } x_1 \in (-a, a)$$

where $t_2^{(0)} = \{0, 0, D_0\}^T$

Then from last equation and (24) we have

$$W^+(x_1) + W^-(x_1) = G^{-1} t_2^{(0)} \quad \text{for } x_1 \in (-a, a) \tag{33}$$

Introducing new function $\mathbf{Y}(z)$ by means the formula

$$W(z) = \mathbf{Y}(z) + \frac{1}{2} G^{-1} t_2^{(0)}, \tag{34}$$

we get from (33):

$$\mathbf{Y}^+(x_1) + \mathbf{Y}^-(x_1) = 0 \quad \text{for } x_1 \in (-a, a). \tag{35}$$

Condition at infinity (28) transforms into the form

$$\mathbf{Y}(z)\big|_{z \to \infty} = \frac{1}{2} G^{-1} \left( t_2^\infty - t_2^{(0)} \right). \tag{36}$$

The solution of the problem (35), (36) takes the form:

$$\mathbf{Y}(z) = \frac{1}{2} G^{-1} \left( t_2^\infty - t_2^{(0)} \right) \frac{z}{\sqrt{z^2 - a^2}} \tag{37}$$

Using further (24), (34) and (37) we get

$$t_2^{(1)}(x_1, 0) = G\left( \mathbf{Y}^+(x_1) + \mathbf{Y}^-(x_1) \right) + t_2^{(0)},$$

Taking into account that $\mathbf{Y}^+(x_1) = \mathbf{Y}^-(x_1) = \mathbf{Y}(x_1)$ for $x_1 \notin (-a, a)$ one obtains

$$t_2^{(1)}(x_1, 0) = 2G \, \mathbf{Y}(x_1) + t_2^{(0)} = \left( t_2^\infty - t_2^{(0)} \right) \frac{z}{\sqrt{z^2 - a^2}} + t_2^{(0)} \quad \text{for } x_1 \notin (-a, a). \tag{38}$$

Equation (38) in the complete form can be written as

$$\sigma_{23}(x_1,0) = \sigma_{32}^{\infty}\frac{z}{\sqrt{z^2-a^2}}, \quad H_{23}(x_1,0) = H_{32}^{\infty}\frac{z}{\sqrt{z^2-a^2}}, \quad D_2(x_1,0) = \left(D_2^{\infty}-D_0\right)\frac{z}{\sqrt{z^2-a^2}} + D_0 \quad (39)$$

The stress and electric displacement intensity factors (SIF) at the point $a$ can be introduced in the following vector form:

$$\mathbf{K} = \lim_{x_1 \to a-0}\sqrt{2\pi(x_1-a)}\,\mathbf{t}_2^{(1)}(x_1,0), \quad (40)$$

where $\mathbf{K} = [K_{3\sigma}, K_{3H}, K_D]^T$

Substituting the expressions (38), (39) into (40) we get

$$K_{3\sigma} = \sqrt{\pi a}\,\sigma_{32}^{\infty}, \quad K_{3H} = \sqrt{\pi a}\,H_{32}^{\infty}, \quad K_D = \sqrt{\pi a}\left(D_2^{\infty}-D_0\right) \quad (41)$$

On the base of (20), (34) and (35) we arrive to the formula

$$\langle \mathbf{u}'(x_1)\rangle = \mathbf{G}^{-1}\left(\mathbf{t}_2^{\infty}-\mathbf{t}_2^{(0)}\right)\frac{x_1}{i\sqrt{a^2-x_1^2}} \quad \text{for } x_1 \in (-a,a), \quad (42)$$

which gives after integration

$$\langle \mathbf{u}(x_1)\rangle = i\mathbf{G}^{-1}\left(\mathbf{t}_2^{\infty}-\mathbf{t}_2^{(0)}\right)\sqrt{a^2-x_1^2} \quad \text{for } x_1 \in (-a,a). \quad (43)$$

Expression (43) in the complete form is the following

$$\begin{Bmatrix}\langle u_3\rangle \\ \langle w_3\rangle \\ \langle \varphi\rangle\end{Bmatrix} = \mathbf{M}\begin{Bmatrix}\sigma_{32}^{\infty} \\ H_{32}^{\infty} \\ D_2^{\infty}-D_0\end{Bmatrix}\sqrt{a^2-x_1^2} \quad (44)$$

where $\mathbf{M} = i\mathbf{G}^{-1}$.

## 5. Electrically limited permeable crack

Consider now another crack model in which the relation between the electric potential and displacement jumps along the crack region is taken into account. We assume now that the cracks filler has the dielectric permittivity

$$\varepsilon_a = \varepsilon_r \varepsilon_0,$$

where $\varepsilon_r$ is the relative dielectric permittivity and $\varepsilon_0 = 8.85\times 10^{-12}\,C/Vm$ is the dielectric constant of a vacuum. We assume also that the crack faces are free of prescribed mechanical loading and electric charges. Moreover, similarly to [12] we consider that the electric field inside the crack can be found as

$$E_a = -\frac{\varphi^+ - \varphi^-}{u_3^+ - u_3^-} \quad \text{for } x_1 \in (-a,a).$$

where the superscripts "+" and "−" indicate the upper and lower faces of the interface, respectively.

Taking into account that $D_2 = \varepsilon_a E_a$ and designating $D_2$ at the crack faces as $D_0$, one arrives to the following electric condition

$$D_0 = -\varepsilon_a\frac{\varphi^+ - \varphi^-}{u_3^+ - u_3^-} = -\varepsilon_a\frac{\langle\varphi\rangle}{\langle u_3\rangle} \quad \text{for } x_1 \in (-a,a) \quad (45)$$

along the crack region. It is worth to be mentioned that we use the same designation for the electric displacement on the crack faces as in the previous section, but in this case $D_0$ is unknown and should be found from Eq. (45).

Taking into account that Eq. (44) remains valid in this case, the Eq. (45) can be written in the following form

$$D_0 = -\varepsilon_a \frac{M_{31}\sigma_{32}^\infty + M_{32}H_{32}^\infty + M_{33}(D_2^\infty - D_0)}{M_{11}\sigma_{32}^\infty + M_{12}H_{32}^\infty + M_{13}(D_2^\infty - D_0)} \quad (46)$$

The equation (46) represents the quadratic equation with respect to the electric displacement $D_0$. After simplification it can be written as

$$D_0^2 - \chi_1 D_0 - \chi_2 = 0 \quad (47)$$

where

$$\chi_1 = \left(M_{11}\sigma_{32}^\infty + M_{12}H_{32}^\infty + M_{13}D_2^\infty - \varepsilon_a\right)/M_{13}, \quad \chi_2 = \varepsilon_a\left(M_{31}\sigma_{32}^\infty + M_{32}H_{32}^\infty + M_{33}D_2^\infty\right)/M_{13}$$

The analysis of the solution of the Eq. (47) shows that one of two its roots is physically unacceptable, therefore, in the following we'll pay our attention only to the physically acceptable root of this equation.

## 6. Numerical illustration and discussion

In this section the main attention will be devoted to the electrically limited permeable crack model, which best reflect the physical peculiarities of the crack deformation. Consider for the analysis the piezoelectric QCs with the following values of the required parameters [15, 16]:
$c_{44}^{(1)} = 5.0\times10^{10}\,Pa$, $e_{15}^{(1)} = -0.318\,C/m^2$, $K_2^{(1)} = 0.3\times10^9\,Pa$, $R_3^{(1)} = 1.2\times10^9\,Pa$, $\hat{e}_{15}^{(1)} = -0.16\,C/m^2$, $\xi_{11}^{(1)} = 8.25\times10^{-11}\,C^2/(Nm^2)$ for the upper material,
$c_{44}^{(2)} = 3.55\times10^{10}\,Pa$, $e_{15}^{(2)} = 17\,C/m^2$, $K_2^{(2)} = 0.15\times10^9\,Pa$, $R_3^{(2)} = 1.765\times10^9\,Pa$, $\hat{e}_{15}^{(2)} = 17\,C/m^2$, $\xi_{11}^{(2)} = 15.1\times10^{-9}\,C^2/(Nm^2)$ for the lower one.
It is worth to mention that the matrix $\mathbf{M}$ in this case is the following

$$\mathbf{M} = \begin{bmatrix} 5.3090\times10^{-11} & -9.8852\times10^{-11} & 0.03892 \\ -9.8852\times10^{-11} & -1.9204\times10^{-9} & -3.2408 \\ 0.03892 & -3.2408 & -5.8781\times10^9 \end{bmatrix}.$$

For the illustration we'll chose $a = 0.01\,m$ and the cases of the external loading given in Table 1. In the last column of this Table the corresponding values of electric flux $D_0$, obtained from the Eq. (47), are given.

**Table 1.** Load cases and corresponding electric flux $D_0$

| Case | $10^{-6}\sigma_{32}^\infty$ [Pa] | $10^{-6}H_{32}^\infty$ [Pa] | $100D_2^\infty$ [C/m²] | $100D_0$ [C/m²] |
|---|---|---|---|---|
| Case 1 | 6.0 | 7.0 | 0.5 | 0.9506 |
| Case 2 | 8.0 | 9.0 | 0.75 | 1.1982 |
| Case 3 | 10.0 | 11.0 | 1.0 | 1.5651 |

It is seen from the Table 1 that rather close values of external loading lead to very different magnitudes of the electric flux through the crack region.

The graphs of phonon, phason and electric quantities along the material interface are given in the Figures 2-4. Particularly phonon shear stress $\sigma_{32}(x_1,0)$ [Pa] at the right crack continuation and the phonon crack faces jump (crack sliding) $\langle u_3(x_1)\rangle$ [m] are shown in Fig. 2 (a) and (b), respectively. Here and in the following figures the solid, dashed and dot lines are drawn for the cases 1, 2 and 3, respectively.

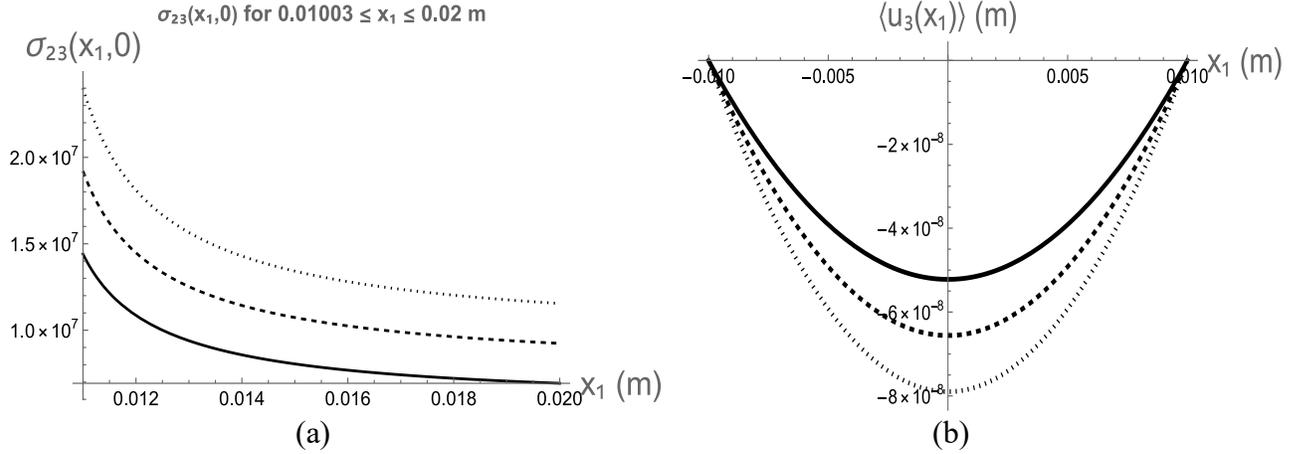

**Fig. 2**. Phonon shear stress at the right crack continuation (a) and the phonon crack faces jump (b) for three cases of the loading

Phason stress $H_{32}(x_1,0)$ [Pa] at the right crack continuation and the phason crack faces jump $\langle w_3(x_1) \rangle$ [m] are shown in Fig. 3 (a) and (b), respectively.

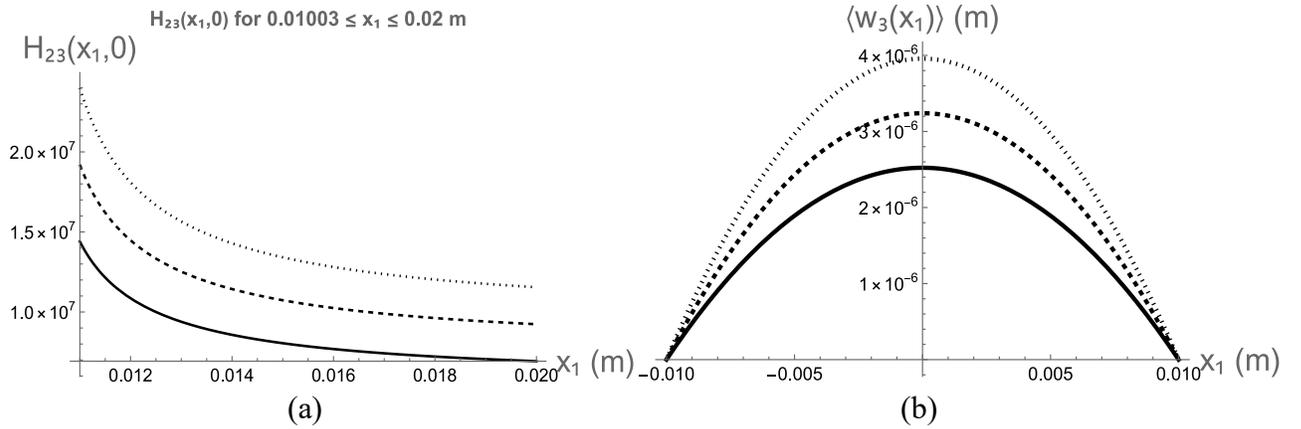

**Fig. 3**. Phason stress at the right crack continuation (a) and the phonon crack faces jump (b)

The electric displacement $D_2(x_1,0)$ [C/m²] at the right crack continuation and the electric field jump over the crack faces $\langle \varphi(x_1) \rangle$ [V] are shown in Fig. 4 (a) and (b), respectively.

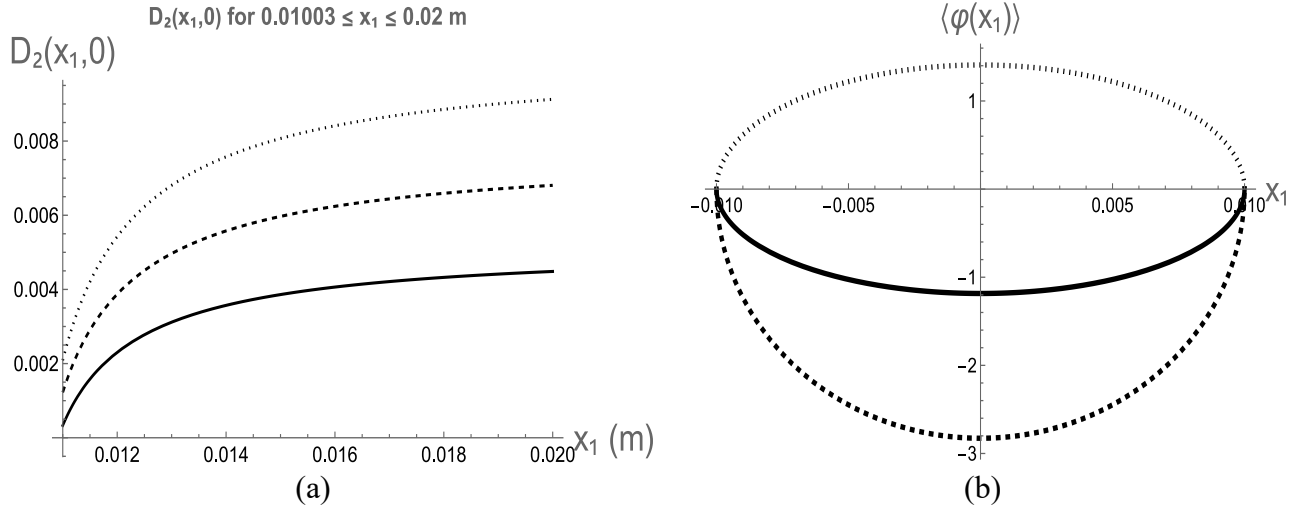

**Fig. 4.** Electric displacement at the right crack continuation and the electric field jump through the crack faces

As it can be seen from the formulas (38), (39) phonon and phason stresses and the electric displacement have the square root singularity at the crack tips. On this reason their graphs in the figures 2(a), 3(a) and 4(a) are drown only in the interval (0.01003 m $<x_1<$ 0.02 m) e.i. on some distance from the crack tip.

The SIFs for three selected cases of the loading were found on the formulas (41) and are presented in Table 2.

**Table 2:** Stress Intensity Factors (SIF) for three selected cases

| Case | $K_{3\sigma}$ [Pa√m] | $K_{3H}$ [Pa√m] | $K_D$ [C $m^{3/2}$] |
|---|---|---|---|
| Case 1 | 1.504×10⁷ | 1.755×10⁷ | 0.02211 |
| Case 2 | 2.005×10⁷ | 2.256×10⁷ | 0.03110 |
| Case 3 | 2.507×10⁷ | 2.757×10⁷ | 0.04010 |

### 7. Conclusions

Mode III interface crack in one-dimentional quasicrystal with piezoelectric effect under anti-plane mechanical and inplane electric loading is considered. Complex function method was used and all electromechanical parameters are presented through vector-functions analytic in the whole complex plane except the crack region by means of equations (20), (24). Electrically impermeable and electrically limited permeable crack models are assumed for analysis. First case leads to the vector Hilbert problem (27) with the condition at infinity (28) and the second one requires additionally the solution of the quadratic equation (47) with respect to the electric flux through the crack region. The case of nonzero prescribed electric displacement on the crack faces is also considered in section 4. For all considered kinds of electrical conditions the analytical formulas for phonon and phason stresses, displacement jumps, electric components along the material interface and also for the correspondisng stress intensity factors at the crack tip are found.

Three variants of the loading given in Table 1 were chosen for the numerical illustration. The corresponding values of electric flux $D_0$ are also given in this table. Other numerical results are summarised in Figures 2–4 and Table 2. Particularly phonon shear stress $\sigma_{32}(x_1,0)$, phason stress

$H_{32}(x_1,0)$ and the electric displacement $D_2(x_1,0)$ along the right crack continuation are given in Figures 2 (a) – 4 (a), respectively. Phonon $\langle u_3(x_1) \rangle$ and phason $\langle w_3(x_1) \rangle$ crack faces displacement jumps are shown in Figures 2 (b) and 3 (b), respectively, and the electric potential jump is given in Fig 4 (b). It follows from the obtained results that along the crack-continuation region beyond the crack, the phonon and phason shear stresses decay proportionally to the inverse square root of the distance from the tip, whereas the electric displacement converges more gradually to its remote value. Besides, rather close values of external loading lead to very different magnitudes of the electric flux through the crack region and both kinds of crack faces displacement and electric potential jumps maintain the square root parabolic behavior along the crack region.